\documentstyle[12pt]{article}
\normalbaselines
\catcode `\@=11
\@addtoreset{equation}{section}

\def\section{\@startsection {section}{1}{\z@}{-3.5ex plus -1ex minus
     -.2ex}{2.3ex plus .2ex}{\normalsize\bf}}
\def\subsection{\@startsection{subsection}{2}{\z@}{-3.25ex plus -1ex minus
 -.2ex}{1.5ex plus .2ex}{\normalsize\bf}}

\def\thebibliography#1{\section*{References\markboth
  {REFERENCES}{REFERENCES}}\list
  {[\arabic{enumi}]}{\settowidth\labelwidth{[#1]}\leftmargin\labelwidth
  \advance\leftmargin\labelsep
  \usecounter{enumi}}
  \def\newblock{\hskip .11em plus .33em minus -.07em}
  \sloppy
  \sfcode`\.=1000\relax}
 
\catcode `\@=12

\skip\footins = 14pt 
\begin{document}
\noindent
MIT-CTP-2783
\hfil September 1998 
\break
-----------------------------------------------------------------------------------------------------------------

\vspace*{2.2cm}
\noindent
{ \bf EFFECTIVE DYNAMICS ON A LINE }\vspace{1.3cm}\\
\noindent
\hspace*{1in}
\begin{minipage}{13cm}
P. Maraner \vspace{0.3cm}\\
\makebox[3mm]Center for Theoretical Physics,\\
\makebox[3mm]Laboratory for Nuclear Science and INFN,\\ 
\makebox[3mm]Massachusetts Institute of Technology, \\
\makebox[3mm]Cambridge, MA 02139-4307, USA
\end{minipage}

\vspace*{0.5cm}

\begin{abstract}
\noindent
The effective classical/quantum dynamics of a particle constrained  
on a closed line embedded in a higher dimensional configuration space 
is analyzed. By considering explicit examples it is shown how different 
reduction mechanisms produce unequivalent dynamical behaviors. 
The relation with a formal treatment of the constraint is discussed.
While classically it is always possible to strictly enforce the 
constraint by setting to zero the energy stored in the motion 
normal to the constraint surface, the quantum description
is far more sensitive to the reduction mechanism.
Not only quantum dynamics is plagued by the usual ambiguities inherent 
to the quantization procedure, but also in some cases the constraint's 
equations do not contain all the necessary information to reconstruct
the effective motion.
\end{abstract}

\noindent
 In this paper we would like to discuss a few aspects of reduction of 
particles motion from a higher dimensional configuration space to
a line. This problem --as the more general one of reducing on a submanifold 
of arbitrary codimension-- appears in physics in the most different contexts.
From the general theory of constrained systems in mathematical physics, to the 
construction of confining devices in solid states and plasma physics,
to the analysis of dynamics around solitonic solutions in non-linear
field theories, to more speculative applications in attempts to 
unification of fundamental interactions. 

\noindent
From the viewpoint of the theory of constrained systems, the 
problem is trivially solvable --up to ambiguities inherent 
to the quantization procedure \cite{HT}.
At the classical level one proceeds by adapting coordinates and imposes 
constraints by freezing motion in directions normal to the constraint 
surface. 
If the system is not subject to forces tangent to the constraint,  
the procedure yields free dynamics. For a line we trivially obtain 
${\cal H}_{cons}={1\over2}{p_\xi}^2$ in the arc-length parameterization $\xi$. 
The quantum mechanical problem is slightly more complicated. We may 
proceed by reducing the classical theory and quantizing the corresponding 
Dirac brackets {\em or} by quantizing the whole theory and imposing 
constraints as functional conditions on quantum states. 
The two strategies turn out to be equivalent, yielding the classically 
expected result plus a {\em quantum gauge connection} ${\cal A}$ and a  
{\em quantum potential} ${\cal Q}$ depending on the quantization 
prescriptions, ${\cal H}_{cons}={1\over2}(-i\partial_\xi-e{\cal A})^2+
{\cal Q}$ \cite{OF}.

\noindent
 On the other hand, from a practical viewpoint,  effective motion 
on a submanifold of the original configuration space is dynamically
produced in a variety of different physical systems. The corresponding
effective dynamics turns in general to be more rich than the one 
obtained by Dirac's algorithm. The character depends strongly on the 
peculiar reduction mechanism. We are going to illustrate this 
--and eventually its relation with the theory of constrained systems-- 
by considering three different mechanisms.
The first one is relevant in a variety of contexts,
its most popular application being perhaps in solid state physics \cite{V}.
The second  one is borrowed from plasma physics \cite{A} and the third one 
form high energy physics \cite{g}. 
We consider a line embedded in an $(n+1)$-dimensional manifold 
${\cal M}$.
In order to retain the maximal amount of  information on the induced dynamics
we close the line into a loop $L$. This makes it possible to consider 
non-trivial gauge interactions in one dimension.
Classical/quantum dynamics on ${\cal M}$ is tentatively described by a 
Lagrangian quadratic in the velocities\footnote{The restriction of 
considering an effective one-dimensional non-relativistic dynamics is 
not essential. The effective motion on a line retains all the features 
of the reduction mechanism, avoiding complications produced by a 
non-trivial intrinsic geometry and the corresponding heaviness in the
notation. Moreover, the reduction mechanisms we consider depend 
only on the spatial structure and not on time, making the 
relativistic and non-relativistic problems essentially analogous.
The generalization to an arbitrary codimension and relativistic 
dynamics is almost straightforward.}
\begin{eqnarray}
{\cal L}={1\over2}g_{ij}{\dot x}^i{\dot x}^j +A_i{\dot x}^i+V
\nonumber
\end{eqnarray}     
Three kinds of interactions act on the particle
\begin{description}
\item{--} a gravitational-like force described by a non-trivial metric 
          $g_{ij}(x)$
\item{--} a magnetic-like force described by a closed antisymmetric two-form
          $B_{ij}$(x) or by the associated vector potential $A_i(x)$
\item{--} a scalar potential $V(x)$
\end{description}
Under appropriate conditions each one of these interactions may produce
an effective one dimensional dynamics. Dimensional reduction produced 
by a nontrivial metric (topology) goes back to the ideas of Kaluza and 
Klein \cite{g}. Variation on this theme are still a key ingredient in 
todays attempts to  unification \cite{OW}. 
Confinement by a magnetic field was first considered by 
Alfv\'en in the fifties and is part of everydays work for plasma physicists
\cite{A}. 
The scalar potential mechanism is used in the construction of 2d quantum 
Hall devices and is relevant in many other applications \cite{V}.   

\noindent
 By reconsidering these mechanisms in a single prospective we show
how different dynamical behaviors are produced by different 
reduction procedures. Of course, all these models have been extensively  
studied in the literature. We concentrate on rather unconventional and 
sometimes unexplored aspects. Our focus is on the geometry of the 
reduction mechanism.

\section[]{\hspace{-4mm}.\hspace{2mm} 
            Reducing by a scalar potential}

\noindent
 The first model we consider is obtained by enforcing the constraint 
by a scalar potential. This mechanism is used in the construction of 
2d quantum Hall devices
and finds interesting application in molecular and chemical physics \cite{V}.
In the context of unification of fundamental interaction it is somehow 
related to the Rubakov-Shaposhnikov model \cite{RS}  
which has recently attracted new attention in the literature \cite{DS,Br}.
The reduction on a line by a potential has been discussed over the years 
by many authors \cite{ma} and presented in a final form by Takagi and 
Tanzawa \cite{TT}.
We consider a particle moving in the Euclidean space ${\cal M}=R^{n+1}$
--we first focus on the case $n=2$--
under the action of a potential $V(x)$ satisfying the following conditions: 
$i)$  $V$ presents a deep minimum in correspondence of a loop $L$ and 
$ii)$ $V$ depends only on the distance from the loop
\begin{eqnarray}
{\cal L}={1\over2}{\dot x}^i{\dot x}^i - V
\end{eqnarray}
The particle experiences  a force attracting it toward the loop $L$. 
No forces act in the tangent direction. A very deep minimum of the potential 
$V(x)$ traps the particle in a narrow neighborhood of the line
producing an effective one dimensional motion. In order to find the 
explicit form of the induced dynamics we proceed by adapting coordinates.
Denoting by ${\bf t}$ the tangent vector and introducing in every 
point an orthonormal frame $\{{\bf n}_1,{\bf n}_2\}$ of the normal space, 
we consider the coordinate transformation $x\rightarrow 
(\xi,\nu^a; a=1,2)$: $\xi$ is the arc-length on $L$ measured from some 
reference point and $\nu_1$, $\nu_2$ are the distances along the geodesics 
leaving the line with velocity ${\bf n}_1$ and ${\bf n}_2$ respectively. 
In this coordinate frame the flat $R^3$ metrics rewrites as
\begin{equation}
g_{ij}=
\left(
\begin{array}{cc}
(1-\kappa_a\nu^a)^{1/2}+\nu^2\tau^2  & \tau\varepsilon_{bc}\nu^c \cr
\tau\varepsilon_{ac}\nu^c &  \delta_{ab}
\end{array}
\right)
\end{equation}
where $\nu=\sqrt{{\nu^1}^2+{\nu^2}^2}$ is the distance from the line, 
$\kappa_a={\bf n}_a\cdot\nabla_{\bf t}{\bf t}$ are the {\em extrinsic
curvatures} of the line and 
$\tau\varepsilon_{ab}={\bf n}_a\cdot \nabla_{\bf t}{\bf n}_b$ is its
{\em generalized torsion}\footnote{When ${\bf n}_1$ and ${\bf n}_2$ are 
chosen as the {\em normal} ${\bf n}$ and the {\em binormal} ${\bf b}$ to the 
line, $\kappa_1$ correspond to the curvature $k$, $\kappa_2$ is zero and 
$\tau$ equals the torsion $t$. In a generic frame $k=\kappa=\sqrt{
{\kappa_1}^2+{\kappa_2}^2}$ and 
$t=\tau-\partial_\xi\chi$, $\chi(\xi)$ being the angle between 
$\{{\bf n}_1,{\bf n}_2\}$ and $\{{\bf n},{\bf b}\}$.}.
Observe that a rotation of the normal frame $\{{\bf n}_1,{\bf n}_2\}$ by 
a point dependent angle $\chi(\xi)$ produce $\tau$ to transform as a 
$SO(2)$ gauge connection, $\tau\rightarrow\tau+\partial_\xi\chi$. 

\noindent
Quantum dynamics is described by the univocally
defined Hamiltonian ${\cal H}=-{1\over2}\partial_i\partial_i+V(\nu)$.
To enforce the constraint we expand $V$ around its minimum keeping only
the quadratic term: $V(\nu)=\nu^2/2\epsilon^2$. ${\cal H}$ expands then 
in a power series in the small parameter $\epsilon$. The first term of the 
expansion is an harmonic oscillator in the normal variables and diverges as
$\epsilon^{-1}$. It represents the fluctuations of the particle when squeezed 
on the constraint surface. The subsequent term is independent on $\epsilon$
and have to be identified with the Hamiltonian describing the effective 
motion along the line. Freezing the normal oscillation in an eigenstate,
ordinary perturbation theory or averaging techniques are used to separate 
the dynamics along the line from the one in the normal directions. The 
procedure  yields \cite{TT}
\begin{eqnarray}
{\cal H}^{V}_{eff}={1\over2}(-i\partial_\xi-e\tau)^2+{\cal Q}^V
\label{HV}
\end{eqnarray} 
The effective dynamics is by no means free. The particle experiences 
a coupling with a gauge field and a univocally defined quantum potential. 
The gauge connection is proportional to the torsion of the line and survives 
in the classical limit. The coupling 
constant $e$ corresponds to the angular momentum stored in the normal 
oscillations and is quantized in integer multiplies of $\hbar$. 
The gauge group $SO(2)$ correspond to the group of the normal bundle of
the line.
Generalizing to an arbitrary codimension $n$ we obtain in fact the gauge group
$SO(n)$.  This is broken in the direct product of orthogonal 
groups of lower dimensionality  if the potential is not completely symmetric
--i.e. does not depend only on the distance from the line \cite{gf}.  
The scalar interaction is of pure quantal nature and is proportional 
to the extrinsic curvature of the line, ${\cal Q}^V= -{\kappa}^2/8$.

\noindent
 Thought not deeply of a geometrical nature, this model 
is substantially equivalent to the Rubakov-Shaposhnikov  
unification scenario which has recently attracted new 
attention \cite{OW,DS,Br}. An interesting feature of the 
model is that it is possible to generate the grand-unified 
group $SO(10)$ in a very natural manner.

\section[]{\hspace{-4mm}.\hspace{2mm} 
            Reducing by a vector potential}

\noindent
 We next consider dimensional reduction produced by a magnetic-like force
$B_{ij}$. This mechanism is commonly employed in plasma physics in confining 
charged particles inside mirror machines and is based on the so called 
{\em guiding center approximation} \cite{A}. First consider a charged particle
moving in a homogeneous magnetic field of strength $B$. Its trajectory is 
an helix of radius $\simeq1/B$ wrapping around a straight field line. 
 The particle performs a fast rotation in the plane normal to the field 
and propagates freely along the magnetic line. The conservation of the 
guiding center position --the center of the circular orbit-- prevents 
the particle from drifting in the directions normal to the line. 
In the strong field limit the rotational motion becomes undetectable and 
the particle behaves as effectively confined on the straight field line. 
Next consider the motion in an inhomogeneous field ${\vec B}(x)$. Provided 
the field is strong enough, the motion still decomposes on three
different energy scales. A fast rotation of radius $\simeq 1/|{\vec B}(x)|$ 
and energy $\simeq |{\vec B(x)}|$, a drift along the field lines with every 
of order one and a very slow drift in the directions normal to the field 
with  energy $\simeq1/|{\vec B}(x)|$. The conservation of the guiding center 
position is in general broken. When the magnetic field norm is a constant 
however, $|{\vec B}(x)|=1/\epsilon$, it is still possible to consider the 
formal limit $\epsilon\rightarrow0$ and trap the particle on a magnetic 
field line.

\noindent
From a geometrical point of view this model is more appealing
than the previous one. An antisymmetric two-form $B_{ij}$ is a quite 
common object in geometry and is generally introduced to describe a 
complex structure on a manifold \cite{Ya}. In that context however $B_{ij}$ is 
required to be non-degenerate while the degenerate directions of the 
magnetic field are precisely the submanifold on which we are reducing 
dynamics. We therefore describe the configuration space of the system 
as a manifold having a somehow mixed real and complex structure. 
We introduce a $2m+1$ dimensional manifold ${\cal M}$ --to start with 
we set $m=1$-- endowed with a Riemannian metric $g_{ij}$ and a closed 
antisymmetric two-form $B_{ij}$ of rank $2m$. We further
assume the norm $\sqrt{B_{ij}B^{ij}}=1/\epsilon$ to be constant. This 
represents a geometry having one {\em real} and $m$ {\em complex} directions.
The real direction closes up in loops $L$. Dynamics on ${\cal M}$ is free in 
the sense of this `half-real/half-complex' geometry
\begin{eqnarray}
{\cal L}={1\over2}g_{ij}{\dot x}^i{\dot x}^j+A_i{\dot x}^i
\end{eqnarray}
where $A_i$ is the vector potential representing the two-form $B_{ij}$. 
We expect the strong field regime to produce an effective one dimensional 
dynamics along a real direction of ${\cal M}$. In order to obtain the explicit
form we proceed again by adapting coordinates. 
We introduce an Euler-Darboux 
frame $x\rightarrow(\xi,\delta_a;a=1,2)$: $\xi$ is the arc-length along 
every magnetic field line while $\delta_1$, $\delta_2$ are coordinates 
bringing $B_{ij}$ in canonical form \cite{St}. Metric and magnetic 
two-form rewrites as 
\begin{equation}
g_{ij}=
\left(
\begin{array}{cc}
1  & a_b \cr
a_a & \gamma_{ab}+a_aa_b
\end{array}
\right)
\ \ \ \ \ \ 
B_{ij}=
\left(
\begin{array}{cc}
0  & 0 \cr
0  &  \varepsilon_{ab}
\end{array}
\right)
\end{equation}
where $\varepsilon_{ab}$ is the completely antisymmetric tensor in two
dimensions and $a_a$, $\gamma_{ab}$ may be  related to the 
the quantities characterizing the foliation of ${\cal M}$ in its real 
directions: the {\em extrinsic curvatures} $\kappa_a$ and the 
{\em generalized torsion} $\tau$ of every line plus the {\em expansion} 
$\theta_{ab}$ and the {\em vorticity} $\omega$ defined as the symmetric 
and antisymmetric part of ${\bf t}\cdot\nabla_{{\bf n}_a}{\bf n}_b$ 
respectively\footnote{The orthonormal frame  $\{{\bf t},{\bf n}_1,
{\bf n}_2\}$ is constructed along every line $L$ as in the previous 
paragraph.}. 

\noindent
In the adapted frame the Hamiltonian operator 
${\cal H}=-{1\over2}g^{ij}(\nabla_i-iA_i)(\nabla_j-iA_j)+\alpha R$  
 --defined up to curvature ambiguities-- expands naturally in powers of 
$\epsilon$.
A quite complex procedure of averaging allows to separate the 
various freedoms in the first few terms of the expansion \cite{Ma}.
The first term is an harmonic oscillator representing the fast rotation 
of the system around the line. As in the previous case it diverges as 
$\epsilon^{-1}$. The second term is of order one and has to be identified
with the effective Hamiltonian describing dynamics on the line. Freezing
the system in a harmonic oscillator eigenstate we obtain   
\begin{eqnarray}
{\cal H}^{\vec A}_{eff}={1\over2}(-i\partial_\xi-e(\tau+\omega))^2-
               e^2\left({1\over4}R+ {1\over4}\kappa^2
               -{1\over2}\theta^2+{1\over2}\omega^2\right)
               +{\cal Q}^{\vec A}
\label{HA}
\end{eqnarray}
Once again the effective dynamics is not free. The particle experiences 
a gauge as well as scalar force. These depend now not only on the 
extrinsic properties of the constraint surface but also on the 
geometry of the foliation of the space ${\cal M}$ in its real
directions. The gauge connection, as an example, is  proportional to the sum 
of the torsion of the line and the vorticity of the foliation along the 
line. It survives in the classical limit.
The coupling constant $e$ corresponds to the energy stored in the 
fast rotation around the line. In the quantum description it
is always different from zero and is quantized in half-integer multiples 
of $\hbar$. The gauge group $U(1)$ corresponds again to the symmetry group 
of the normal double of the constraint surface. In the present model, however, 
the normal space  carries a complex structure, so that the generalization to 
an arbitrary codimension $2m$ produces the gauge group $U(1)\times SU(m)$. 
The scalar interaction consists in a part
surviving in the classical limit plus a pure quantum contribution 
${\cal Q}^{\vec A}=(\alpha-{1\over16})R-{1\over16}\kappa^2-{3\over16}\theta^2
+{1\over8}\omega^2$. Both depend on the scalar curvature $R$ of ${\cal M}$,
the extrinsic curvature $\kappa$, the expansion rate  
$\theta=\sqrt{\theta_{ab}\theta_{ab}}$ and the vorticity $\omega$
of the foliation.

\section[]{\hspace{-4mm}.\hspace{2mm} 
            Reducing by a metric (topology)}

\noindent
Dimensional reduction produced by a metric --generated indeed by a 
non-trivial topological background-- may be obtained by means of the
Kaluza-Klein mechanism \cite{g}. Though this procedure is not 
directly assimilable to the reduction on a subspace of the original 
configuration space, we briefly review it here for comparison with 
other models. In order to obtain an effective one dimensional dynamics 
we start with a configuration space ${\cal M}=L\times\Sigma$ given by 
the direct 
product of a loop $L$ times a compact manifold $\Sigma$ with a group
of isometries $G$. The original Kaluza-Klein mechanism considers
$\Sigma$ to be the circle $S^1$, $G=U(1)$. For now we restrict our attention 
to this simple case. Dynamics on ${\cal M}$ is assumed 
to be free
\begin{eqnarray}
{\cal L}={1\over2}g_{ij}{\dot x}^i{\dot x}^j
\end{eqnarray}
The basic idea is that when the size of $\Sigma$ is shrunk to zero the 
motion along the directions of the compact manifold becomes undetectable 
and we are left with an effective one-dimensional configurations space.   
 To find out the explicit form of the effective dynamics it is convenient
to parameterize the metric in the form
\begin{equation}
g_{ij}=
\left(
\begin{array}{cc}
1+ a^2/2v  &  a/2v \cr
a/2v & 1/2v
\end{array}
\right)
\label{m3}
\end{equation}
The first coordinate $\xi$ is again the arc-length along $L$ while 
the second coordinate $\sigma$ parameterizes the circle $S^1$. Observe 
that a $\xi$ dependent translation of $\sigma$, $(\xi,\sigma)\rightarrow
(\xi,\sigma-\chi(\xi))$ produces $a$ to transform as a vector potential,
$a\rightarrow a+\partial_\xi\chi$. The `covariance' of (\ref{m3}) under this 
transformation indicates that the embedding of $L$ in ${\cal M}$ is not 
relevant for this model.

\noindent
Quantum dynamics on ${\cal M}$ is described the Hamiltonian operator 
${\cal H}=-{1\over2}g^{ij}\nabla_i\nabla_j+\alpha R$ --defined 
up to curvature ambiguities.
Dimensional reduction is implemented by expanding $a$ and $v$ in 
a Fourier series in the compact direction and retaining only the 
zero order harmonic --basically by averaging over $\sigma$.
This produces the effective Hamiltonian
\begin{eqnarray}
{\cal H}^{g_{ij}}_{eff}={1\over2}(-i\partial_\xi-e\langle a \rangle)^2+
                                 e^2\langle v \rangle +{\cal Q}^{g_{ij}}
\label{Hg}
\end{eqnarray}
where $\langle a\rangle=\int_{S^1}a(\xi,\sigma)d\sigma/
\int_{S^1}d\sigma$ and $\langle v\rangle=\int_{S^1}v(\xi,\sigma)d\sigma/
\int_{S^1}d\sigma$ indicates the average of $a$ and $v$ in the compact
direction.
The particle experiences again a gauge and a scalar interaction. Since 
the effective configuration space is not a submanifold embedded in 
${\cal M}$ these are not directly connected to any extrinsic geometrical
quantity. The gauge force survives in the classical limit.
The coupling constant $e$ corresponds to the momentum 
stored in the compact direction and is again quantized in integer multiples    
of $\hbar$. The gauge group $U(1)$ coincides to the group of isometries 
of the circle $S^1$. Generalizing to an arbitrary codimension --with an
appropriate ansatz on the metric-- yields in fact the gauge group G. 
The scalar interaction consists of the classical potential $\langle 
v\rangle$ plus a pure quantum contribution ${\cal Q}^{g_{ij}}$ depending 
on the average  of first and second derivatives of $a$ and $v$ over $\sigma$. 

\section[]{\hspace{-4mm}.\hspace{2mm} 
            Discussion: reducing vs. constraining }

The first two models we considered --the dynamical reduction form a higher
dimensional configuration space to a line produced by a scalar and 
a vector potential-- represent two different {\em physical realizations}
of the same constrained system. Unequivalent dynamical behaviors are
generated on the line: the analytic form of the induced gauge connection 
and of the scalar potential are different, the induced gauge group is 
different and even the coupling constant $e$ is quantized in a different way.
It is therefore natural to wonder about the relation between these 
models and the formal treatment of the constraint according to the 
methods of analytical mechanics. 

\noindent
We first consider the classical
theory, setting $-i\partial_\xi\rightarrow p_\xi$ and 
${\cal Q}^{V,{\vec A},g_{ij}}\rightarrow 0$ in (\ref{HV}), (\ref{HA})
and (\ref{Hg}). The difference between the effective Hamiltonians 
${\cal H}^{V}_{eff}$, 
${\cal H}^{\vec A}_{eff}$, ${\cal H}^{g_{ij}}_{eff}$ and the 
result expected from a formal analysis, ${\cal H}_{cons}=
{1\over2}{p_\xi}^2$, is then given by the terms proportional to 
the effective coupling constant $e$ and to its square $e^2$. In all
the models we discussed, $e$ keeps somehow track of the energy stored in 
the motion normal to the constraint's surface. 
A strict enforcement of the constraint requires 
as a compatibility condition the momenta in the directions normal to the 
constraint to be zero. That is $e=0$. On the other hand,
 from a physical viewpoint, the constraint is enforced when it is 
impossible to detect  deviations of the motion form the 
line,  so that the 
system is actually free to move in the directions normal to the line
on length scales less than the experimental resolution.
The track of this hidden motion is kept by the induced interactions. 
 There is no contradiction between a {\em physical
realization} and a {\em formal treatment} of the constraint, once 
we remember that we are working in the hypothesis $e=0$.

\noindent
Much more subtle is the situation in quantum mechanics. The uncertainty
principle forbids to set simultaneously to zero a coordinate and its 
conjugate momentum, so that a strict enforcement of the constraint 
is {\em a priori} impossible. This corresponds to the well known 
fact that we are not allowed to impose simultaneously all the constraints 
as functional conditions on the quantum states {\em or}, equivalently,
that the quantization of Dirac's brackets is defined up to terms of 
order $\hbar^2$. A track of the mechanism producing the dimensional 
reduction on the effective constrained dynamics seems therefore to be 
unavoidable.
Not knowing anything about the reduction mechanism
--we suppose to start the analysis from the classical theory
plus its constraints-- it is extremely appealing that the ambiguities 
inherent to the construction of the quantum theory allow us the freedom
to reproduce different situations. There are nevertheless two facts 
that have to be remarked. The first one is that extremely different 
dynamical behaviors may be produced by choosing different 
quantization procedure / reduction mechanisms. 
As an example, we may be tempted to quantize the motion on the line as free,
while in our first model the quantum potential ${\cal Q}^{V}$ 
attracts and possibly confines particles around points of strong curvature.
 This information, nevertheless, is  somehow contained in the 
constraint's equations and is therefore predictable to some extent.
 The second and perhaps more remarkable point that emerges form 
our analysis, is that in some circumstances the effective 
constrained dynamics depends on informations which are not contained 
in the constraint's equations. In our second model, the effective 
Hamiltonian (\ref{HA}) depends on {\em expansion} and {\em vorticity} 
of the foliation of ${\cal M}$ in its real directions. These quantities 
are not computable starting from the constraint's  equations.

\noindent

\section*{\hspace{4mm} Acknowledgments}
 It is a real pleasure to thank Domenico Seminara for helpful 
discussions and George Lavrelashvili for bringing ref.\cite{RS} 
to my attention.

\end{document}